\documentclass[%
 aip,
 jmp,%
 amsmath,amssymb,
 reprint,%
]{revtex4-1}

\usepackage{graphicx}
\usepackage{dcolumn}
\usepackage{bm}

\def\capfigure{figure}

\long\def\@makecaption#1#2{%

  \vskip\abovecaptionskip

  \ifx\@captype\capfigure

      \centering #1~--~#2 \par

  \else

      #1~--~#2 \par

  \fi

  \vskip\belowcaptionskip}
\begin{document}

\preprint{APS/123-QED}

\title{Magnetic and magnetotransport properties of Bi$_2$Se$_3$ thin films doped by Eu}

\author{B.A. Aronzon}
\thanks{Corresponding author \emph{e}-mail address: aronzon@mail.ru}
\affiliation{National Research Center "Kurchatov institute", 123182 Moscow, Russia}
\affiliation{P.N. Lebedev Physical Institute RAS, 119991 Moscow, Russia}
\author{L.N. Oveshnikov}
\affiliation{National Research Center "Kurchatov institute", 123182 Moscow, Russia}
\affiliation{P.N. Lebedev Physical Institute RAS, 119991 Moscow, Russia}
\author{V.A. Prudkoglyad}
\affiliation{P.N. Lebedev Physical Institute RAS, 119991 Moscow, Russia}
\author{Yu.G. Selivanov}
\affiliation{P.N. Lebedev Physical Institute RAS, 119991 Moscow, Russia}
\author{E.G. Chizhevskii}
\affiliation{P.N. Lebedev Physical Institute RAS, 119991 Moscow, Russia}
\author{K.I. Kugel}
\affiliation{Institute for Theoretical and Applied Electrodynamics RAS, 125412 Moscow, Russia}
\author{I.A. Karateev}
\affiliation{National Research Center "Kurchatov institute", 123182 Moscow, Russia}
\author{A.L. Vasiliev}
\affiliation{National Research Center "Kurchatov institute", 123182 Moscow, Russia}
\author{E. L$\ddot{\mathrm a}$hderanta}
\affiliation{Lappeenranta University of Technology, 53850 Lappeenranta, Finland}


\date{\today}

\begin{abstract}
Structural, magnetic and magnetotransport properties of (Bi$_{1-x}$Eu$_x$)$_2$Se$_3$ thin films have been studied experimentally as a function of Eu content. The films were synthesized by MBE. It is demonstrated that Eu distribution is not uniform, it enter quint-layers forming inside them plain (pancake-like) areas containing Eu atoms, which sizes and concentration increase with the growth of Eu content. Positive magnetoresistance related to the weak antilocalization was observed up to 15K. The antilocalization was not followed by weak localization as theory predicts for nontrivial topological states. Surprisingly, the features of antilocalization were seen even at Eu content $x$ $=$ 0.21. With the increase of Eu content the transition to ferromagnetic state occurs at $x$ about 0.1 and with the Curie temperature $\approx$ 8K, that rises up to 64K for $x$ $=$ 0.21. At temperatures above 1-2 K, the dephasing length is proportional to $T^{-1/2}$ indicating the dominant contribution of inelastic \emph{e-e} scattering into electron phase breaking. However, at low temperatures the dephasing length saturates, that could be due to the scattering on magnetic ions.
\end{abstract}

\maketitle


\section{Introduction}
3D topological insulators are novel materials and an innovative research field in the condensed matter physics. The distinct point is the study of properties of these materials with magnetic dopants, so called magnetic topological insulators. The investigations in that direction reveal unusual quantum effects such as the quantum anomalous Hall effect and quantum magnetoelectric effect \cite{a1,a2,a3}. Although the properties of gapless interface states in the presence of magnetic impurities have been studied for more than 30 years \cite{Kus}, detailed investigation and explanation of the specific features of electron transport and of magnetic phenomena in 3D magnetic topological insulators remain an unresolved scientific problem even nowadays, being the subject of great interest \cite{a4,a5}. We used Eu as magnetic impurity ($J$ $=$ $7/2$) with the aim to study the effect of magnetic impurities on electron transport in 3D topological insulators.

We report on magnetic properties of Eu-doped Bi$_2$Se$_3$ and effect of magnetic impurities on electron transport, including weak antilocalization effect within the wide range of magnetic fields (up to 16T) and temperatures (0.3-30 K) as a function of Eu content. Magnetic measurements were performed with SQUID magnetometer at temperatures in the range 2-70 K in magnetic fields up to 7T.

\section{Samples}
\begin{table}[t]
\begin{center}
\begin{tabular}{|c|*{10}{c|}}
\hline
\textbf{Sample} & $\emph{\textbf{x}}$ & $\emph{\textbf{d}}$, & \mbox{\boldmath$\mu$}, & $\emph{\textbf{n}}$, \\
  & & nm & cm$^2$/$(V\cdot s$) & 10$^{13}$ cm$^{-2}$ \\
\hline
 737 & 0.009 & 28 & $\approx$ 780 & 3.6 \\
\hline
 733 & 0.038 & 24 & $\approx$ 150 & 6.2 \\
\hline
 735 & 0.07 & 26 & $\approx$ 270 & 9.4 \\
\hline
 736 & 0.13 & 23 & $\approx$ 90 & 28 \\
\hline
 719 & 0.21 & 60 & $\approx$ 80 & 8.3 \\
\hline
\end{tabular}
\caption{Sample parameters: Eu content $x$; film thickness $d$; carrier mobility $\mu$; carrier concentration $n$.}
\end{center}
\end{table}
The main parameters of studied samples are listed in the Table 1. Eu-doped Bi$_2$Se$_3$ thin films with the thicknesses of 20 to 60 nm were grown on (111) BaF$_2$ substrates in a MBE apparatus EP-1201 with a pressure 1$\cdot$10$^{-10}$ Torr, specially designed for the growth of chalcogenide films  \cite{a6}. For simplicity, we will refer to our doped films as ternary compounds, although structural data suggests highly inhomogeneous distribution of Eu impurities. (Bi$_{1-x}$Eu$_x$)$_2$Se$_3$ thin films were deposited using standard effusion cells for high purity elemental Se, Eu and binary Bi$_2$Se$_3$ compound. The beam equivalent pressure (BEP) of molecular/atomic beams was measured by Bayard-Alpert ion gage that swings into the substrate position.  The cells temperatures for Se (130$^\circ$C) and Bi$_2$Se$_3$ (495$^\circ$C) were kept constant which resulted in Se-rich conditions with a BEP flux ratio Se/Bi$_2$Se$_3$ of 2$:$1 and provided layer growth rate of $\sim$ 0.25 nm/min. The Eu content was regulated by varying the evaporation temperature of the Eu effusion cell. Heating Eu cell within the 320-430 $^\circ$C range resulted in the increase of Eu content $x$ from 0.002 to 0.21 in the grown films. Concentration $x$ of Eu in the (Bi$_{1-x}$Eu$_x$)$_2$Se$_3$ layers was deduced from the growth rates of the Bi$_2$Se$_3$ \cite{a7}  and EuSe reference layers, grown on (111) BaF$_2$ substrates.

For Eu cell temperatures up to 405$^\circ$C, streaky Reflection High-Energy Electron Diffraction (RHEED) pattern was observed at the very beginning of the growth, and it got brighter as the substrate temperature reached 300$^\circ$C. For Eu cell temperatures above 410$^\circ$C the streaks became more faint and diffusive, indicating that the increase of Eu percentage lowers crystalline quality. X-Ray Diffraction study revealed that up to the $x$ $\sim$ 0.13 Eu-doped Bi$_2$Se$_3$ thin films have the same rhombohedral crystal structure as binary Bi$_2$Se$_3$, while (\emph{003l}) peak family gets broader and less intensive indicating lower crystalline quality of the films with increasing Eu concentration. For samples with $x$ $>$ 0.13 X-Ray Diffraction spectra show contributions from different crystal structures.

With the growth of Eu content $x$, High-Angle Annular Dark-Field (HAADF) Scanning Transmission Electron Microscopy (STEM) imaging and Energy Dispersive X-ray spectroscopy analysis revealed Eu-enriched nanoclusters with increasing lateral dimensions (10 to 30 nm) and concentration (2$\cdot$10$^{16}$ to 2$\cdot$10$^{17}$ cm$^{-3}$).

To protect surface of the films from exposure to the atmosphere and to stabilize their properties, films were \emph{in situ} covered by the 30-40 nm thick layer of amorphous selenium.

\section{Structural investigation}
Structural investigation by HAADF STEM was performed to reveal the microstructure of the samples.
\begin{figure}[h!]
\centering
\includegraphics[width=8cm,clip]{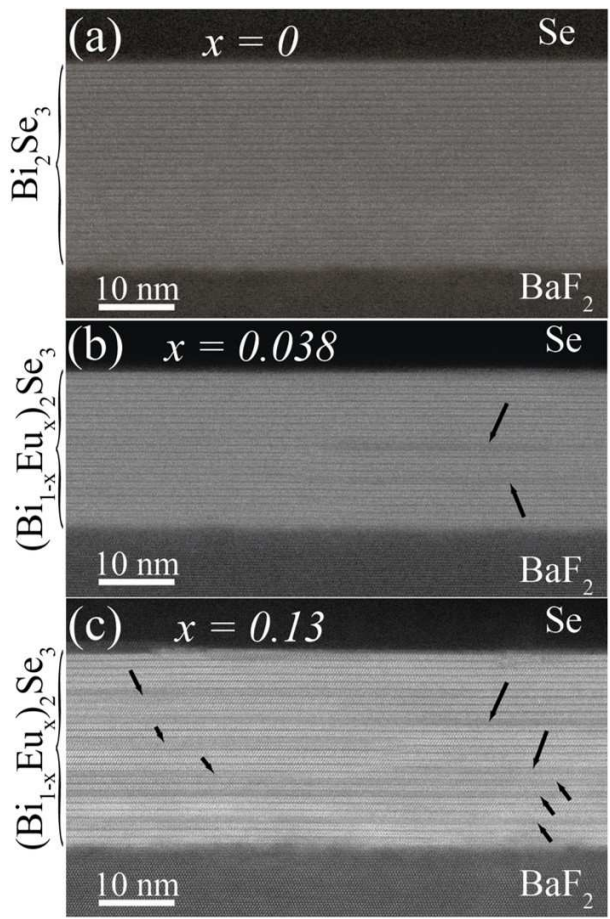}
\caption{HAADF STEM images of Bi$_2$Se$_3$ with different Eu content $x$: (a) - pure film, (b) - 0.038, (c) - 0.13. Black arrows indicate the pancake-like defects.}
\label{fig-1}       
\end{figure}

The cross-section specimens were prepared by standard focus ion beam (FIB) procedure in Helios (FEI, USA) dual beam instrument. The low magnification HAADF STEM images of samples with $x$ $=$ 0, 0.038 and 0.13 are presented in Fig. 1(a-c), respectively. The image of pure Bi$_2$Se$_3$ sample (see Fig.1(a)) demonstrates the absence of structural defects in the film. The Eu doping leads to the formation of flat pancake-like defects, shown by black arrows in Fig. 1(b,c). The results of Energy Dispersive X-Ray spectroscopy and Electron Energy Loss spectroscopy (not presented here) suggests the presence of Eu only in these pancake-like regions. The increase of Eu content leads to the growth of lateral dimensions of these regions from $\sim$10 to $\sim$30 nm and their density from $\sim$2$\cdot$10$^{16}$ to $\sim$2$\cdot$10$^{17}$ cm$^{-3}$, and that can be seen in Fig. 1(b,c).

\section{Magnetoresistance measurements}
\begin{figure}[h!]
\centering
\includegraphics[width=8cm,clip]{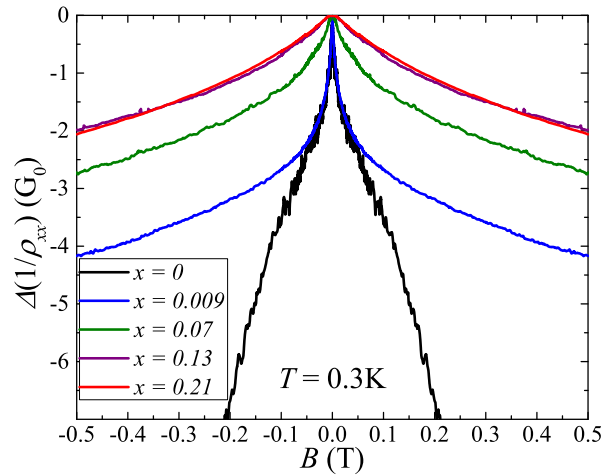}
\caption{(colour online) Magnetoconductivity curves for samples with various Eu content $x$ including pure Bi$_2$Se$_3$ film \cite{a6} at weak fields. The curves demonstrate the antilocalization features. With the increase of Eu content the sharpness of antilocalization peak decreases.  }
\label{fig-1}       
\end{figure}

\begin{figure}[h!]
\centering
\includegraphics[width=8cm,clip]{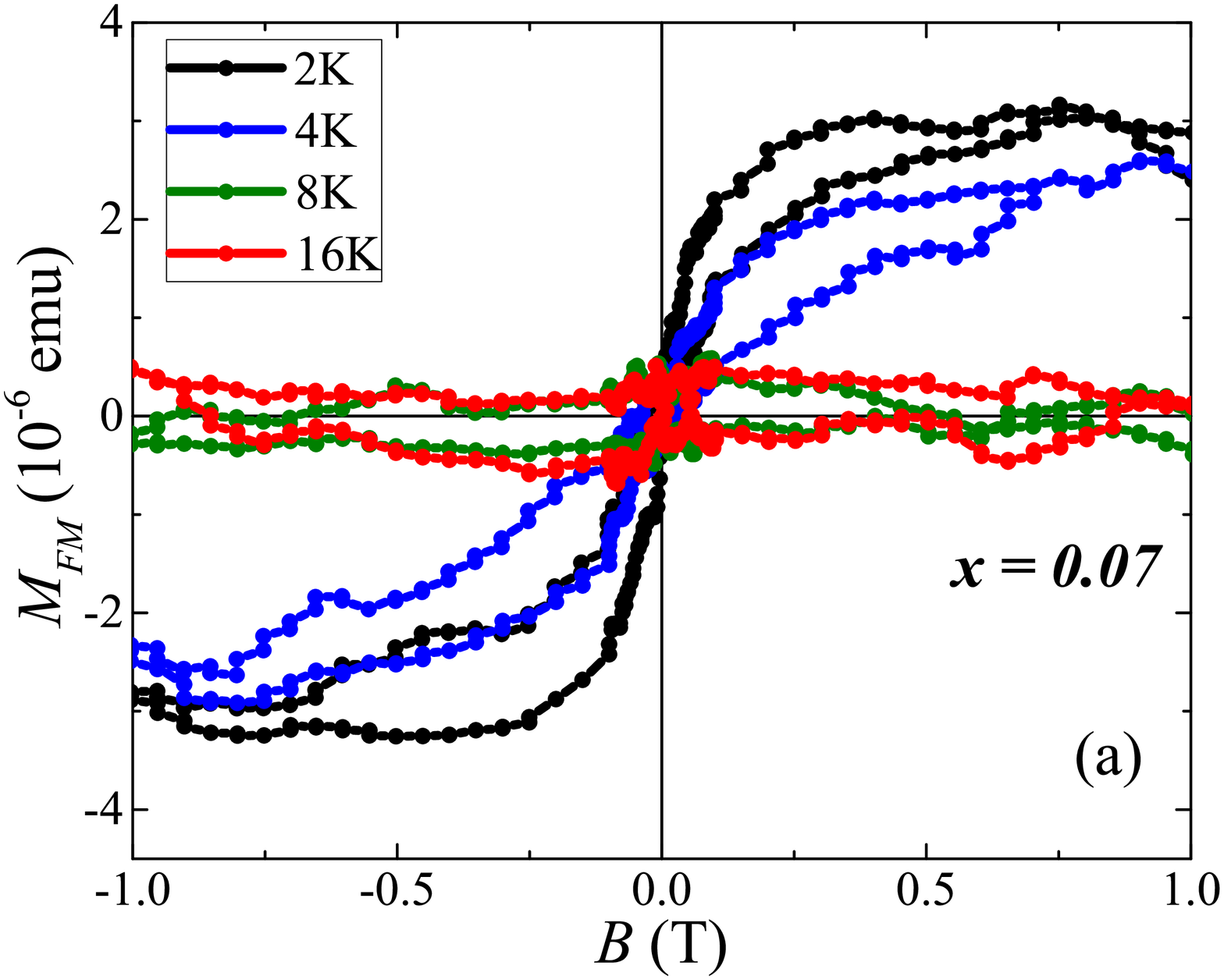}
\includegraphics[width=8cm,clip]{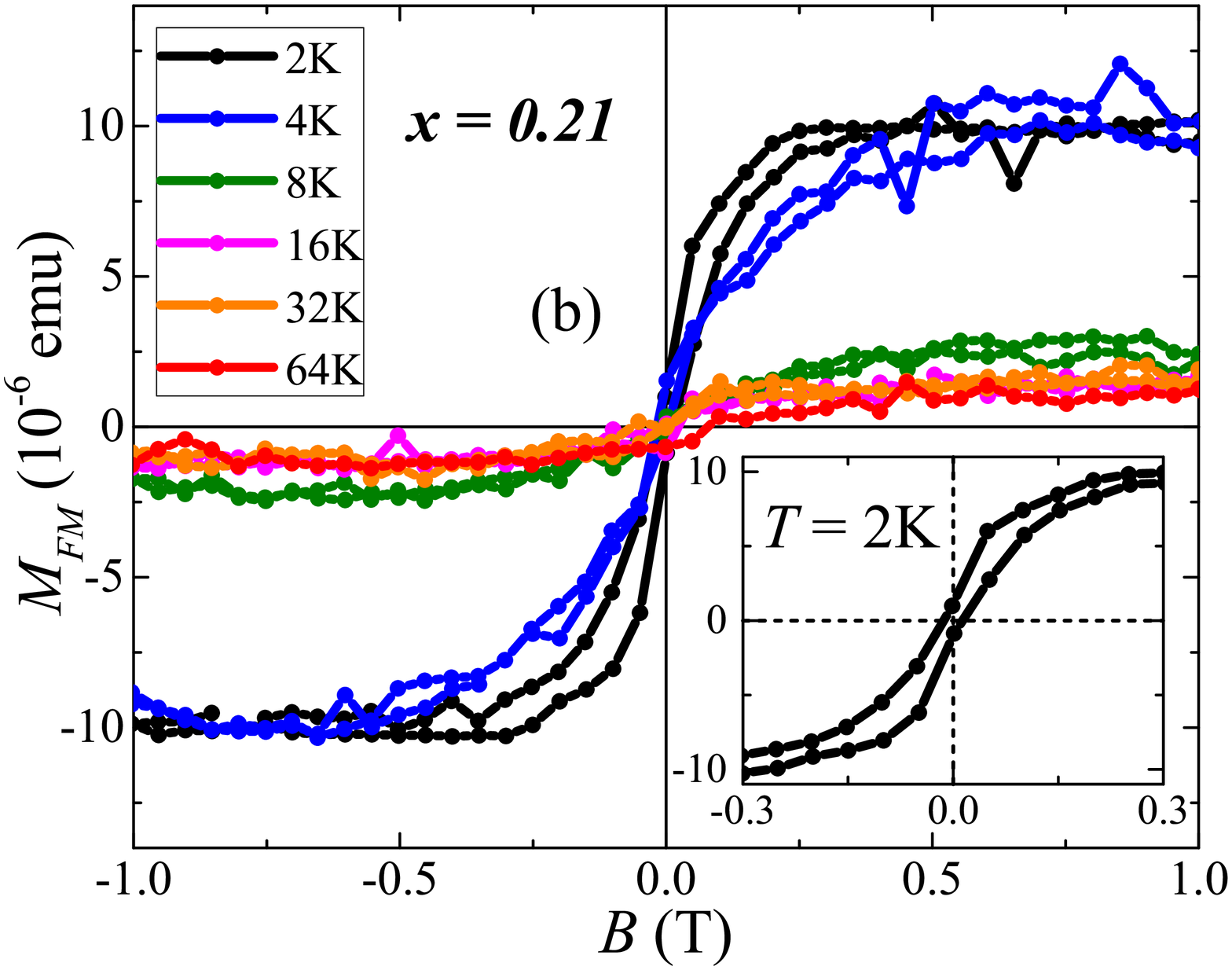}
\includegraphics[width=8cm,clip]{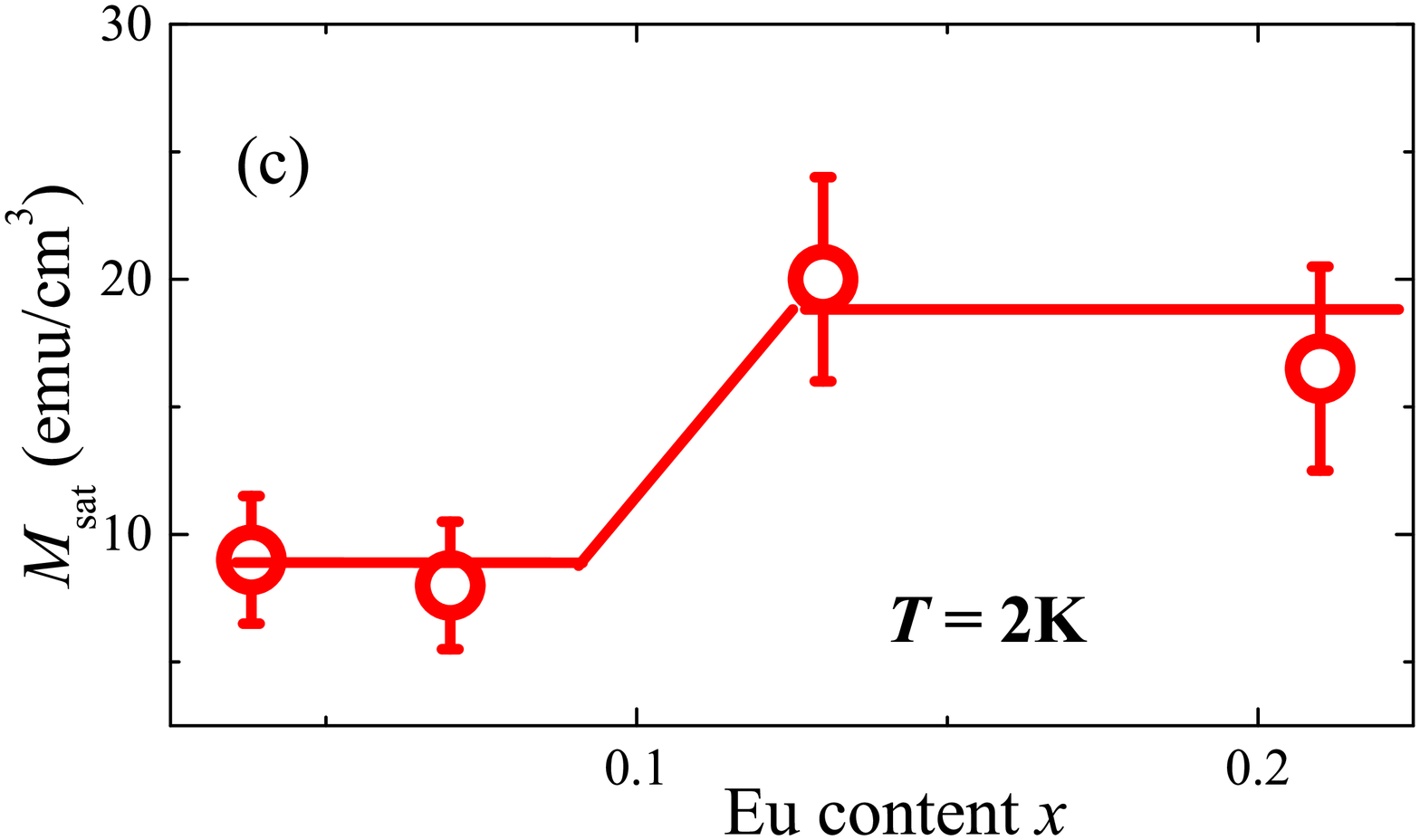}
\caption{(colour online) The magnetization dependence on magnetic field for samples with Eu content of $x$ $=$ 0.07 (a) and $x$ $=$ 0.21 (b) at various temperatures. Inset on (b) shows magnified image of low-temperature hysteresis for corresponding sample. (c) Saturation magnetization measured at temperature 2K versus Eu content.}
\label{fig-1}       
\end{figure}

The magnetoconductivity $\Delta(1/\rho_{xx})(B)$ of an analogous undoped Bi$_2$Se$_3$ films was studied previously \cite{a6}. It was shown that the negative magnetoconductivity (weak antilocalization), as well as the Shubnikov-de Haas (SdH) oscillations for pure samples are determined only by the magnetic field component perpendicular to the film plane. The obtained experimental results suggested that the studied films exhibit two-dimensional topologically protected electron states. Moreover, the estimated contribution of these states to the total conductivity was the dominant one. Unlike the pure Bi$_2$Se$_3$ films, the carrier mobility in studied samples (see Table 1) is lower, and no pronounced SdH oscillations were observed, thus we were not able to distinguish bulk and surface contributions to the total conductivity explicitly. However, the measurements at various directions of magnetic fields demonstrated that contributions of bulk states and surface states are different and the surface states contribution is essential. It manifests in substantial magnetoresistance (MR) anisotropy. Namely, for Eu-doped films the MR in magnetic field perpendicular to the sample plane is more than 2 times higher than for in-plane field orientation. While for in-plane geometry, the MR data measured with field perpendicular and parallel to charge current almost coincide. It should be noted, that for conventional bulk conductors MR should be the same for magnetic field oriented perpendicular to the current, both for in-plane and perpendicular to the plane orientations.

The theory predicts that for nontrivial topological states one should observe the negative magnetoconductivity $\Delta(1/\rho_{xx})(B)$ related to antilocalization correction to conductivity, which is not followed by the positive magnetoconductivity related to weak localization \cite{a8,a9}. Such behavior was observed in our measurements, which results are presented in Fig. 2, being the hint and illustration of nontrivial topology of conduction states in studied samples. This behavior is strikingly different from the usual behavior of systems with trivial, non-topological states.

\section{Magnetic properties}
SQUID magnetic measurements show diamagnetic signal from the substrate while the sample demonstrate both paramagnetic and weak ferromagnetic contributions, which were obtained by subtracting the diamagnetic signal. The paramagnetic signal saturates at low temperatures and at high fields above 4-5 T. Ferromagnetic contribution saturates at fields less than 1T, it can be extracted by subtraction of linear signal. Corresponding contribution for samples with $x$ $=$ 0.07 and 0.21 is presented in Fig. 3(a,b). For samples with $x$ $<$ 0.1, the ferromagnetic signal is weak and originates from magnetic moments of isolated Eu-containing clusters, while for samples with $x$ $=$ 0.13 and $x$ $=$ 0.21, it seems that the long-range ferromagnetic state is established. The dependence of the magnetic moments at saturation on Eu content is presented in Fig. 3(c). It should be mentioned that the ferromagnetic magnetization in this samples is weak compared to the diamagnetic signal related to the sample substrate. That is due to the great difference in thicknesses of Bi$_2$Se$_3$ film and of sample substrate. As a result, the errors of obtained values of (Bi$_{1-x}$Eu$_x$)$_2$Se$_3$ film magnetic moment are about 20$\%$ as it is seen in Fig. 3(c). Nevertheless, the strong changes of saturation magnetization occur at $x$ above 0.1.  The Curie temperature is about 8K for $x$ $=$ 0.13 and for sample with $x$ $=$ 0.21 it is about 64K.

\begin{figure}[h!]
\centering
\includegraphics[width=8cm,clip]{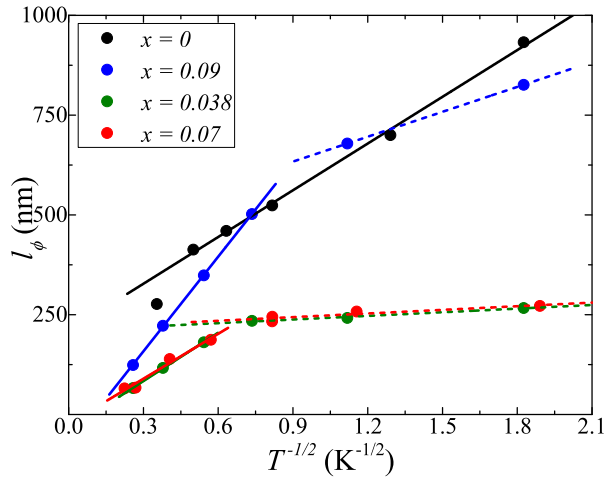}
\caption{(colour online) The temperature dependence of dephasing length for samples with various Eu content $x$. In pure sample \cite{a6} and at temperatures higher than 1-2 K for Eu-doped samples  $l_\phi\propto T^{-1/2}$ that corresponds to phase breaking by inelastic \emph{e-e} scattering. It is clearly seen that dephasing length dependence saturates (tend to saturate) at temperature below 1-2 K for all Eu doped films. }
\label{fig-1}       
\end{figure}

\section{Discussion}
The main feature of the non-trivial topological surface states is the protection of electrons from the back-scattering. This protection is conserved by time reversal symmetry and is related to strong spin-orbit interaction, which leads to a strong coupling of electron spin with its quasi-momentum (spin-momentum locking). As a result, the electron $k$-vector is keeping its direction until spin will be scattered. To scatter spin the magnetic moments are needed and they are provided in our samples by Eu ions. At high concentration magnetic impurities should break the spin-momentum locking, however we have observed some features of weak antilocalization even at $x$ $=$ 0.21, when the long range ferromagnetic state is established. Although antilocalization effect is observed on heavily doped samples, Eu impurities substantially affect the electron transport of (Bi$_{1-x}$Eu$_x$)$_2$Se$_3$ films both in weak and strong magnetic fields.

The magnetoconductivity $\Delta(1/\rho_{xx})(B)$ peak related to weak antilocalization was observed at low magnetic fields for all studied samples (see Fig. 2). With the increase of Eu content the antilocalization peak becomes wider and its amplitude decreases. Corresponding magnetoconductivity was studied using conventional Hikami-Larkin-Nagaoka (HLN) expression for two-dimensional systems:
\begin{displaymath}
\frac{\Delta(1/\rho_{xx}(B))}{G_0}=\alpha \left [ \psi \left ( \frac{1}{2}+\frac{\hbar}{4el_\phi^2 B} \right )-\ln \left ( \frac{\hbar}{4el_\phi^2 B} \right ) \right ],    (1)
\end{displaymath}
where $\psi$ - digamma function, $G_0$ $=$ $e^2/(2\pi^2\hbar)$ ($e$ - electron charge, $\hbar$ - Planck constant), $l_{\phi}$ - dephasing length, $\alpha$ - prefactor. Using HLN expression (1) for fitting of the experimental data we've obtained the temperature dependence of $l_{\phi}$ shown in Fig. 4. As one can see, the depahsing length for undoped sample is $l_\phi\propto T^{-1/2}$ in the whole temperature range while for Eu-doped samples it is valid only for temperatures above $\approx$ 1-2 K. At lower temperatures $l_{\phi}$ saturates. The dependence  $l_\phi\propto T^{-1/2}$ is usually associated with the inelastic \emph{e-e} scattering as a main phase breaking mechanism. The $l_{\phi}$ saturation means that at $T$ $<$ 1-2 K an additional mechanism dominate for the phase breaking and it is natural to suggest that this mechanism is the scattering by magnetic impurities. We would like to mention that in other papers concerning the properties of 3D topological insulators with magnetic impurities the temperature dependence of $l_{\phi}$ was not discussed.

\section{Conclusion}
Thus, we present results on structural, magnetic and magnetotransport properties of (Bi$_{1-x}$Eu$_x$)$_2$Se$_3$ thin films at various Eu content $x$.  The films were grown on (111) BaF$_2$ substrates by MBE technology using standard effusion cells for high purity elemental Se, Eu and binary Bi$_2$Se$_3$ compound. Unlike previous papers \cite{a3,a4,a8,a9} that suggested uniform distribution of magnetic impurities over a sample volume, the Transmission Electron Microscopy images demonstrated that Eu distribution in our case is not uniform. Inside quint-layers there are Eu-enriched pancake-like areas, which sizes and concentration increase with growth of Eu content. We observed the ferromagnetic transition at $x$ $\approx$ 0.13 with the Curie temperature about 8K, while at $x$ $\approx$ 0.21 it reaches 64K. At weak fields, magnetoresistance exhibits the features of weak antilocalization which is not followed by weak localization in accordance with the theoretical prediction for non-trivial topological states. Antilocalization was observed up to 15K and even at high Eu content, above $x$ $=$ 0.21, at which ferromagnetic state is established. For Eu-doped samples with $x$ $>$ 0.01 the dephasing length at $T$ $>$ 1-2 K is determined by inelastic \emph{e-e} scattering, while at low temperature $l_{\phi}(T)$ saturates. We believe that it is due to the contribution of magnetic impurities scattering into electron phase breaking.\\

\begin{acknowledgments}
The work was partly supported by RSF (grant \#17-12-01345). The resource centre of NRC "Kurchatov institute" is acknowledged for the usage of their equipment and help.
\end{acknowledgments}

\end{document}